\documentclass[12pt]{article}
\usepackage[english]{babel}
\usepackage{latexsym}
\usepackage{amsfonts}
\begin{document}
\begin{center}
{\Large \bf A non-standard matter distribution in the RS1 model
and the coupling constant of the radion}\\

\vspace{4mm}

Mikhail N.~Smolyakov, Igor P.~Volobuev\\

\vspace{4mm} Skobeltsyn Institute of Nuclear Physics, Moscow State
University,\\ Vorob'evy Gory, 119992 Moscow, Russia\\
\end{center}

\begin{abstract}
In the zero mode approximation we solve exactly the equations of
motion for linearized gravity in the Randall-Sundrum model with a
non-standard distribution of matter in the neighbourhood of the
negative tension brane. It is shown that the form of this
distribution can strongly affect the coupling of the radion to
matter. We believe that such a situation can arise in models with
a realistic mechanisms of matter localization.
\end{abstract}

\section{Introduction}
The Kaluza-Klein hypothesis has been discussed in theoretical
physics for three quarters of a century. In accordance with this
hypothesis, space-time may have extra dimensions, which are
unobservable for certain reasons. The explanation of this
unobservability, which was put forward in the original papers by
Kaluza and Klein, implies that the extra dimensions are
compactified and have a very small size of the order of the
Planck's length $l_{Pl} = 1/M_{Pl}$.

In 1983 Rubakov and Shaposhnikov put forward a new scenario for
Kaluza-Klein theories, which was based on the idea of localization
of fields on a domain wall \cite{RSH}. They have also proposed an
ansatz for multidimensional metric,  which is compatible with this
hypothesis \cite{RSH1}.

In the last years there appeared indications that scenarios of
this type can arise in the theory of strings
\cite{string,string1,string2,string3} (see Ref. \cite{Ant01} for a
review and references). In this approach our three spatial
dimensions are supposed to be realized as a three-dimensional
hypersurface embedded into a mutidimensional space-time. Such
hypersurfaces are called 3-branes, or just branes. The main goal
of such scenarios was to find a solution to the hierarchy problem.
It was solved either due to the sufficiently large characteristic
size of extra dimensions \cite{ADD}, or due to exponential warp
factor appearing in the metric \cite{RS1}. In both approaches
gravity in multidimensional space-time becomes "strong"\ not at
the energies of the order of $10^{19}\,$GeV, but at much lower
energies, maybe of the order of $1 \div 10\,$TeV. An attractive
feature of these models is that they predict new effects which can
be observed at the coming collider experiments.

In paper \cite{RS1} an exact solution for a system of two branes
interacting with gravity in a five-dimensional space-time $E$ was
found. This model is called the Randall-Sundrum model (usually
abbreviated as RS1 model), and it is widely discussed in the
literature (see Refs. \cite{Rub01,Kubyshin} for reviews and
references). Let us denote the coordinates by $ \{ x^M\} \equiv
\{x^{\mu},y\}$, $M= 0,1,2,3,4, \, \mu=0,1,2,3$, the coordinate
$x^4 \equiv y$ parameterizing the fifth dimension. It forms the
orbifold $S^{1}/Z_{2}$, which is realised as the circle of the
circumference $2R$ with points $y$ and $-y$ identified.
Correspondingly, we have the usual periodicity condition in
space-time $E$, which identifies points $(x, y)$ and $(x, y +
2nR)$, and the metric $g_{MN}$ satisfies the orbifold symmetry
conditions
\begin{eqnarray}
\label{orbifoldsym}
 g_{\mu \nu}(x,- y)=  g_{\mu \nu}(x,  y), \\
 \nonumber
  g_{\mu 4}(x,- y)= - g_{\mu 4}(x,  y), \\ \nonumber
   g_{44}(x,- y)=  g_{44}(x,  y).
\end{eqnarray}
The branes are located at the fixed points of the orbifold, $y=0$
and $y=R$.

The action of the model is
\begin{equation}\label{actionRS}
 S = S_g + S_1 + S_2,
\end{equation}
where $S_g$, $S_1$ and $S_2$ are given by
\begin{eqnarray}\label{actionsRS}
S_g&=& \frac{1}{16 \pi \hat G} \int_E
\left(R-\Lambda\right)\sqrt{-g}\, d^{4}x dy,\\ \nonumber
 S_1&=& V_1 \int_E \sqrt{-\tilde g} \delta(y) d^{4}x dy,\\ \nonumber
 S_2&=& V_2 \int_E \sqrt{-\tilde g}  \delta(y-R) d^{4}x dy.
\end{eqnarray}
{Here $\tilde g_{\mu\nu}$ is the induced metric on the branes and
the subscripts 1 and 2 label the branes.} We also note that the
signature of the metric $g_{MN}$ is chosen to be $(-,+,+,+,+)$.

The Randall-Sundrum solution for the  metric is {given by}
\begin{equation}\label{metricrs}
ds^2=  g_{MN} d{x}^M d{x}^N = \gamma_{\mu\nu} {dx^\mu dx^\nu} +
  dy^2,
\end{equation}
where $\gamma_{\mu\nu}=e^{2\sigma(y)}\eta_{\mu\nu}$,
$\eta_{\mu\nu}$ is the Minkowski metric and {the function}
$\sigma(y) = -k|y|$ in the interval $-R \leq y \leq R$. The
parameter  $k$ is positive and has the dimension of mass, the
parameters $\Lambda$ and $ V_{1,2}$ are {related to it as
follows:} $$ \Lambda = -12 k^2, \quad V_1 =-V_2= -\frac{3k}{4\pi
\hat G}. $$ We see that brane~1 has a positive energy density,
whereas brane~2 has a negative one. The function $\sigma$ has the
properties
\begin{equation}\label{sigma}
  \partial_4 \sigma = -k\, sign(y), \quad \frac{\partial^2 \sigma}{\partial
  {y}^2} =-2k(\delta(y) - \delta(y-R)) \equiv  -2k\tilde \delta .
\end{equation}
Here and in the sequel $\partial_4 \equiv \frac{\partial}{\partial
y}$.

We denote $\hat \kappa = \sqrt{16 \pi \hat G}$, where $\hat G$ is
the five-dimensional gravitational constant, and parameterize the
metric $g_{MN}$ as
\begin{equation}\label{metricpar}
  g_{MN} = \gamma_{MN} + \hat \kappa h_{MN},
\end{equation}
$h_{MN}$ being the metric fluctuations. Substituting this
parameterization into (\ref{actionRS}) and retaining the terms of
the zero order in $\hat \kappa$, we get the second variation
action of this model  \cite{BKSV}. It  is invariant under the
gauge transformations
\begin{eqnarray}\label{gaugetrRS}
h'_{MN}(x,y) = h_{MN}(x,y) -(\nabla_M\xi_N(x,y) +
\nabla_N\xi_M(x,y) ),
\end{eqnarray}
where $\nabla_M$ is the covariant derivative with respect to the
background metric $\gamma_{MN}$, and the functions $\xi_N(x,y)$
satisfy the orbifold symmetry conditions
\begin{eqnarray}\label{orbifoldsym1}
\xi^{\mu}\left(x,-y\right)&=&\xi^{\mu}\left(x,y\right),\\
\nonumber \xi^{4}\left(x,-y\right)&=&-\xi^{4}\left(x,y\right).
\nonumber
\end{eqnarray}
With the help of these gauge transformations we can impose the
gauge
\begin{equation}\label{unitgauge}
h_{\mu4} =0, \, h_{44} = h_{44}(x) \equiv \phi (x),
\end{equation}
which  will be called the {\it unitary gauge} (see \cite{BKSV}).
We would like to emphasize once again that the branes remain
straight in this gauge, i.e. we {\it do not} use the bent-brane
formulation, which allegedly destroys the structure of the model,
i.e. the orbifold symmetry under the  reflection $-y
\leftrightarrow y$ (this problem was discussed in \cite{AIMVV}).

In the linear approximation interaction with matter looks like
\begin{equation}\label{interaction}
 \frac{\hat \kappa}{2} \int h^{MN}(x,y) T_{MN}\sqrt{- \gamma} d^{4}xdy,
\end{equation}
where $ T_{MN}$ is the energy-momentum tensor of the matter: $$
 T_{MN} = 2\frac{\delta L}{\delta  \gamma^{MN}} -
  \gamma_{MN}L.
$$ Thus, we are considering perturbations about the background,
for which $T_{MN}=0$.

We will examine the $h_{\mu\nu}$-components of the metric
fluctuations $h_{MN}$, since all four-dimensional physical effects
can be described in terms of this field. Obviously, the unitary
gauge conditions (\ref{unitgauge}) do not fix the gauge of this
field. In fact, after imposing these gauge conditions there remain
gauge transformations of the form
\begin{equation}\label{remgaugetr}
   \xi_\mu = e^{2\sigma}\epsilon_\mu(x),
\end{equation}
which change the longitudinal components of the field
$h_{\mu\nu}$. Nevertheless, it turns out that it is convenient to
solve the equations of motion for linearized gravity in the
unitary gauge and then to  choose an appropriate gauge  in our
four-dimensional world on the brane. We will use the de Donder
gauge for the field $h_{\mu\nu}$  on the brane, which corresponds
to the choice of  {\it harmonic coordinates}.
\section{Equations of motion}
The equations of motion  for different components of the metric
fluctuations in the unitary gauge take the form (see
\cite{SVHEP,SV}):

 1) $\mu\nu$-component
\begin{eqnarray}\label{mu-nu}
 & &\frac{1}{2}\left(\partial_\rho \partial^\rho h_{\mu\nu}-
\partial_\mu \partial^\rho
h_{\rho\nu}-\partial_\nu \partial^\rho h_{\rho\mu} +
\frac{\partial^2 h_{\mu\nu}}{\partial {x^4}^2}\right)-
2k^2h_{\mu\nu}+\frac{1}{2}\partial_\mu
\partial_\nu\tilde h+ \\
\nonumber &+& \frac{1}{2}\partial_\mu \partial_\nu \phi+
\frac{1}{2} \gamma_{\mu\nu}\left(\partial^\rho
\partial^\sigma h_{\rho\sigma}-\partial_\rho \partial^\rho \tilde
h - \frac{\partial^2 \tilde h}{\partial {x^4}^2}-4\partial_4
\sigma
\partial_4 \tilde h
 - \partial_\rho \partial^\rho \phi + 12 k^2 \phi\right)+\\ \nonumber
&+& \left[2k  h_{\mu\nu} - 3k\gamma_{\mu\nu}\phi \right]\tilde
\delta = -\frac{\hat \kappa}{2} T_{\mu\nu},
\end{eqnarray}

 2) $\mu 4$-component,
\begin{equation}\label{mu-4}
\partial_4 ( \partial_\mu \tilde h - \partial^\nu  h_{\mu\nu})-
3\partial_4 \sigma \partial_\mu \phi = -\hat\kappa T_{\mu 4},
\end{equation}
which plays the role of a constraint,

 3) $4 4$-component
\begin{equation}\label{4-4}
\frac{1}{2}(\partial^\mu \partial^\nu  h_{\mu\nu} - \partial_\mu
\partial^\mu \tilde h ) - \frac{3}{2}\partial_4 \sigma \partial_4 \tilde h
+ 6 k^2 \phi =-\frac{\hat\kappa}{2} T_{4 4},
\end{equation}
with $T_{MN}$ being the energy-momentum tensor of the matter and
$\tilde h=\gamma^{\mu\nu}h_{\mu\nu}$. In what follows, we will
also use an auxiliary equation, which is obtained by multiplying
the equation for $44$-component by 2 and subtracting it from the
contracted equation for $\mu\nu$-component. This equation contains
$\tilde h$ and $\phi$ only and has the form:
\begin{equation}\label{contracted-44}
\frac{\partial^2 \tilde h}{\partial {x^4}^2}  + 2\partial_4 \sigma
\partial_4 \tilde h
  -8k^2 \phi+ 8k \phi \tilde \delta + \partial_\mu \partial^\mu \phi =
\frac{\hat \kappa}{3} \left(T_{\mu}^{\mu}-2T_{4 4}\right).
\end{equation}

For example, if $ T_{MN}=0,$ the physical degrees of freedom of
the model can be extracted by the substitution \cite{BKSV}
\begin{equation}\label{substitution}
 h_{\mu\nu} =  b_{\mu\nu} + \gamma_{\mu\nu}(\sigma - c)\phi +
 \frac{1}{2k^2} \left(\sigma - c +\frac{1}{2} +
 \frac{c}{2}e^{-2\sigma}\right) \partial_\mu \partial_\nu \phi.
\end{equation}
with $c=\frac{kR}{e^{2kR}-1}$. It turns out that the field
$b_{\mu\nu}(x^{\mu},y)$ describes the massless graviton and
massive Kaluza-Klein spin-2 fields, whereas $\phi (x)$ describes a
scalar field called the radion.

However, the situation is rather different, when there is matter
on the branes. The cases, in which matter is located on the
branes, i.e. the energy-momentum tensor is of the form $T_{\mu\nu}
= t_{\mu\nu}(x)\delta(y)$ or $T_{\mu\nu} =
t_{\mu\nu}(x)\delta(y-R)$ (and $T_{\mu 4}=0$, $T_{4 4}=0$) were
discussed in detail in \cite{SVHEP,SV}. It is well known that if
we live on the negative tension brane (at $y=R$), the contribution
of the radion is $e^{2kR}$ times stronger than the contribution of
the massless graviton \cite{SVHEP,SV}. It means that in the case
of the massless radion scalar gravity is realized on brane~2, and
it is necessary to have a mechanism for generating the radion
mass, for example, the Goldberger-Wise mechanism \cite{wise} to
make gravity in the zero mode approximation tensor, i.e. to
suppress the contribution of the scalar radion component. But we
will show below, that this is not the only solution for this
problem.
\section{A simple example}
In the original formulation of the Randall-Sundrum model the
mechanism of localization of fields is not taken into account. The
branes are treated as infinitely thin objects, and it is assumed
that energy-momentum tensor has $\delta(y)$-like profile in the
extra coordinate (matter is located on the brane only).
Nevertheless in some papers (see, for example,
\cite{GS,GRS,Mido2,Oda,GogS}) considering models with extra
dimensions, energy-momentum tensors with dependence on the extra
coordinates were used for constructing non-trivial background
metric. One can also recall models with universal extra
dimensions, in which physical fields can propagate in the extra
dimension \cite{Appelquist}, and fat brane scenarios
\cite{fat1,fat2}. Thus, the existence of energy-momentum tensors
of this type seems to be quite reasonable. Here we consider some
simple examples, which are sufficient to show attractive features
of the non-delta-like localization of matter.

Let us choose as an example the following form of the
energy-momentum tensor
\begin{eqnarray}\label{tensor1}
T_{\mu\nu}&=&t_{\mu\nu}(x)\frac{e^{-2\sigma}c}{R},\\
\label{tensor3} T_{\mu 4}&\equiv& 0,\\ \label{tensor2}
T_{44}&=&t_{\mu\nu}(x)\gamma^{\mu\nu}e^{-2\sigma}\frac{c}{R}\left(\sigma
+e^{2kR}c-\epsilon c\right),
\end{eqnarray}
where $\epsilon$ is some arbitrary constant, which will be defined
later. Here $t_{\mu\nu}(x)$ is the energy-momentum tensor of
matter on the brane, it depends on the four-dimensional
coordinates $x$ only. One can see that the matter is localized
near the negative tension brane at $y=R$. We would like to note
that the energy-momentum tensor of the form (\ref{tensor1}),
(\ref{tensor3}) and (\ref{tensor2}) satisfies the covariant energy
conservation law, which has the form
\begin{equation}\label{econl}
\nabla^{N}T_{MN}=0,
\end{equation}
and the function of localization in (\ref{tensor1}) is normalized
to unity. Constructions similar to that used in this paper, but in
the case of six dimensions, were utilized in \cite{GM1,GM2}. The
explanation of such constructions depends on the method of
localization, and this issue will not be discussed in this paper.

The substitution, which allows one to decouple the equations
(\ref{mu-nu}), (\ref{mu-4}), (\ref{4-4}), (\ref{contracted-44})
with $T_{MN}$ of the form (\ref{tensor1}), (\ref{tensor3}),
(\ref{tensor2}) looks like
\begin{eqnarray}\label{substitution1}
b_{\mu\nu} =  u_{\mu\nu} + \frac{1}{2k^2}
\left(\frac{ce^{2kR}}{\epsilon}-e^{-2\sigma}\left[\frac{ce^{2kR}}{2\epsilon}
+\frac{1}{8\epsilon}\right]-\sigma
e^{-2\sigma}\frac{1}{2\epsilon}\right)
\partial_\mu \partial_\nu \phi,
\end{eqnarray}
where $b_{\mu\nu}$ was defined in (\ref{substitution}) and
$u_{\mu\nu}(x^{\mu},y)$ describes the massless graviton and
massive Kaluza-Klein spin-2 fields. Substituting
(\ref{substitution}) and (\ref{substitution1}) into equations
(\ref{mu-4}), (\ref{4-4}) and (\ref{contracted-44}), we get:
\begin{eqnarray}\label{mu4TT}
\partial_4(e^{-2\sigma}(\partial^\nu u_{\mu\nu}-\partial_\mu
u))=0,
\end{eqnarray}
\begin{eqnarray}\label{44TT}
e^{-4\sigma}(\partial^\mu\partial^\nu u_{\mu\nu}-\Box u) -
3\partial_4\sigma\partial_4(e^{-2\sigma}u)+e^{-2\sigma}
\frac{3ce^{2kR}}{\epsilon}\Box\phi-\\ \nonumber
-\frac{3}{\epsilon}e^{-4\sigma} \left(\sigma+\tilde c
e^{2kR}\right)\Box\phi=-\hat\kappa t\frac{c}{R}e^{-4\sigma}
\left(\sigma+\tilde c e^{2kR}\right),
\end{eqnarray}
\begin{eqnarray}\label{contTT}
\partial_4(e^{2\sigma}\partial_4(e^{-2\sigma}u))+\frac{1}
{\epsilon}\Box\phi e^{-2\sigma}\left(1-2\left(\sigma+\tilde c
e^{2kR}\right)\right)=\\ \nonumber=\frac{\hat\kappa
c}{3R}te^{-2\sigma}\left(1-2\left(\sigma+\tilde c
e^{2kR}\right)\right),
\end{eqnarray}
where $u= \eta^{\mu\nu}u_{\mu\nu}$, $t= \eta^{\mu\nu}t_{\mu\nu}$,
$\partial^{\mu}=\eta^{\mu\nu}\partial_{\nu}$,
$\Box=\eta^{\mu\nu}\partial_{\mu}\partial_{\nu}$ and $$\tilde
c=c\left(1-\epsilon e^{-2kR}\right)$$

Let us consider the Fourier expansion of all terms of  equation
(\ref{contTT}) with respect to coordinate $y$. Since the term with
the derivative $\partial_4$ has no zero mode, this equation
implies that
\begin{equation}\label{eqphi}
\Box \phi=\frac{\hat\kappa\epsilon c}{3R}t,
\end{equation}
\begin{equation}\label{equ}
\partial_4(e^{-2\sigma}u)=0.
\end{equation}
Now let us consider $\mu\nu$-equation. It is well known that the
field $u_{\mu\nu}$ in the presence of matter is a combination of
zero and massive modes, whose eigenfunctions are orthogonal
\cite{BKSV}. In particular, the zero mode can be represented as
$u^0_{\mu\nu}=e^{2\sigma}\alpha_{\mu\nu}$, where $\alpha_{\mu\nu}$
depends on $x$ only. It also means that with the help of the
residual gauge transformations (\ref{remgaugetr}) it is possible
to impose the gauge condition
\begin{eqnarray}\label{gaugeu}
\partial^{\nu}u^m_{\mu\nu}=0, \\ \nonumber
u_{\mu}^{m\mu}=0,
\end{eqnarray}
on the massive modes $u^m_{\mu\nu}$ and the  de Donder gauge
condition on the zero mode
\begin{equation}\label{dedonder2}
\partial^\nu\left(\alpha_{\mu\nu}-\frac{1}{2}\eta_{\mu\nu}
\alpha\right)=0.
\end{equation}
Having imposed this gauge, we are still left with residual gauge
transformation
\begin{equation}\label{ostatgauge}
\xi_\mu = e^{2\sigma}\epsilon_\mu(x), \quad \Box\epsilon_{\mu}=0.
\end{equation}
The gauge transformations with $\xi_\mu$ satisfying these
conditions are important for determining the number of degrees of
freedom of the massless graviton. It follows from equation
(\ref{44TT}) that
\begin{equation}\label{boxu}
\Box \alpha=2\frac{\hat\kappa e^{2kR}c^{2}}{R}t.
\end{equation}
Substituting (\ref{substitution}) and (\ref{substitution1}) into
equation (\ref{mu-nu}) with condition (\ref{eqphi}) and passing to
gauge (\ref{gaugeu}), (\ref{dedonder2}) we get
\begin{eqnarray}\label{munuTT}
\frac{1}{2}\Box(\alpha_{\mu\nu}-\frac{1}{2}\eta_{\mu\nu}\alpha)
 + \frac{1}{2}e^{-2\sigma}\Box
u^m_{\mu\nu}+\frac{1}{2}\partial_4
\partial_4 u^m_{\mu\nu}-2k^2u^m_{\mu\nu}-\partial_4
\partial_4\sigma u^m_{\mu\nu}= \\ \nonumber
=-\frac{\hat\kappa c}{2R}t_{\mu\nu}e^{-2\sigma}+\frac{\hat\kappa
c}{6R}\left(2ce^{2kR}-e^{-2\sigma}\right)\left(\frac{\partial_\mu\partial_\nu}{\Box}-
\eta_{\mu\nu}\right)t
\end{eqnarray}
We are going to calculate the equations of motion in the zero mode
approximation. Thus, we have to find an equation for the field
$\alpha_{\mu\nu}$. If we multiply equation (\ref{munuTT}) by
$e^{2\sigma}$, integrate it over $y$ and take into account the
orthonormality condition for the wave functions of the modes
\cite{BKSV}, we get
\begin{equation}\label{alpha}
\Box (\alpha_{\mu\nu}-\frac{1}{2}\eta_{\mu\nu}\alpha)
=-2\frac{\hat\kappa c^{2}e^{2kR}}{R}t_{\mu\nu}.
\end{equation}
The equation for massive modes takes the form
\begin{eqnarray}
\frac{1}{2}e^{-2\sigma}\Box u^m_{\mu\nu}+\frac{1}{2}\partial_4
\partial_4 u^m_{\mu\nu}-2k^2u^m_{\mu\nu}-\partial_4
\partial_4\sigma u^m_{\mu\nu}= \\ \nonumber
=-\frac{\hat\kappa
c}{2R}t_{\mu\nu}\left(e^{-2\sigma}-2ce^{2kR}\right)+\frac{\hat\kappa
c}{6R}\left(2ce^{2kR}-e^{-2\sigma}\right)\left(\frac{\partial_\mu\partial_\nu}{\Box}-
\eta_{\mu\nu}\right)t
\end{eqnarray}

Now we are ready to find equations of motion for the zero mode
part of $h_{\mu\nu}$. Since the matter is localized near the
brane, we will calculate the equation for $h_{\mu\nu}|_{y=R}$. It
is easy to see that $h_{\mu\nu}|_{y=R}$ (\ref{substitution}),
(\ref{substitution1}) does not satisfy the de Donder gauge
condition. The residual gauge transformations (\ref{ostatgauge})
are not sufficient to pass to this gauge. But since we consider
only the effective theory on brane~2 (at $y=R$), we can drop the
condition $\Box\epsilon_{\mu}=0$, which fixes the gauge for the
field $h_{\mu\nu}$. Then we can pass to the  de Donder gauge
condition for the field $h_{\mu\nu}$ {\it on brane~2} with the
help of these gauge functions. Making these transformations
(analogously to what was made in \cite{SVHEP,SV}), we get in the
zero mode approximation
\begin{eqnarray}\label{subst0}
\left(h_{\mu\nu}-\frac{1}{2}\eta_{\mu\nu}h\right)|_{y=R}=\\
\nonumber = e^{-2kR}\left(\alpha_{\mu\nu}-
\frac{1}{2}\eta_{\mu\nu}\alpha\right)-2e^{-2kR}(-kR
-c)\left(\eta_{\mu\nu}-\frac{\partial_{\mu}\partial_{\nu}}{\Box}\right)\phi.
\end{eqnarray}

An important point is that these equations are written in the
coordinates $\{x^{\mu}\}$, which are {\it Galilean on brane~1}
(not on brane~2) and are inappropriate for studying physical
effects on brane~2 (we recall that coordinates are called
Galilean, if $g_{\mu \nu} = diag(-1,1,1,1)$ \cite{LL}). It is
necessary to pass to Galilean coordinates on brane~2 to get a
correct result. This problem was discussed in detail in papers
\cite{BKSV,SVHEP,SV}. In Galilean coordinates on brane~2 equation
(\ref{subst0}) looks like
\begin{eqnarray}\label{subst1}
\left(h_{\mu\nu}-\frac{1}{2}\eta_{\mu\nu}h\right)|_{y=R}=\\
\nonumber =e^{-2kR}\left(\alpha_{\mu\nu}-
\frac{1}{2}\eta_{\mu\nu}\alpha\right)-2(-kR
-c)\left(\eta_{\mu\nu}-\frac{\partial_{\mu}\partial_{\nu}}{\Box}\right)\phi,
\end{eqnarray}
and equations (\ref{eqphi}), (\ref{alpha}) take the form
\begin{equation}\label{eqphi1}
\Box \phi=\frac{\hat\kappa\epsilon c}{3R}t,
\end{equation}
\begin{equation}\label{alpha1}
\Box (\alpha_{\mu\nu}-\frac{1}{2}\eta_{\mu\nu}\alpha)
=-2\frac{\hat\kappa c^{2}e^{4kR}}{R}t_{\mu\nu}.
\end{equation}
Thus, we can get
\begin{eqnarray}\label{eqbrane}
\Box\left(h_{\mu\nu}-\frac{1}{2}\eta_{\mu\nu}h\right)|_{y=R}=
-\frac{2\hat\kappa c^{2}
e^{2kR}}{R}\left(t_{\mu\nu}-\frac{\epsilon}{3}
\left(\eta_{\mu\nu}-\frac{\partial_{\mu}\partial_{\nu}}{\Box}\right)t\right),
\end{eqnarray}
where $t_{\mu\nu}$ is the energy-momentum tensor of matter in the
coordinates, which are Galilean at $y=R$. One can see, that this
equation coincides with equation for the fluctuations of metric in
the linearized Brans-Dicke theory, which looks like
\begin{eqnarray}\label{BDeq}
\Box \left(\delta  g _{\mu\nu}-\frac{1}{2}\eta_{\mu\nu}\delta
g\right)=-16\pi G \left(t_{\mu\nu}-\frac{1}{2\omega+3}\left(
\eta_{\mu\nu}-\frac{\partial_\mu\partial_\nu}{\Box}\right)t\right),
\end{eqnarray}
where $\omega$ is the BD-parameter and $G$ is the gravitational
constant. Thus, comparing equations (\ref{eqbrane}) and
(\ref{BDeq}), we get
\begin{equation}\label{Gconst}
G=\hat G\frac{2c^{2}e^{2kR}}{R},
\end{equation}
\begin{equation}\label{Omegaconst}
\omega=\frac{3(1-\epsilon)}{2\epsilon}.
\end{equation}
It is easy to see, that in the the Randall-Sundrum model with the
$\delta$-function-like localization of matter on the brane there
is a factor $e^{2kR}$ instead of $\epsilon$ in (\ref{eqbrane})
(compare with the analogous formula in \cite{SVHEP,SV}).

From the recent experimental data (see, for example, \cite{Chiba})
we know, that $\omega > 3500$. It means, that
\begin{equation}
\epsilon<4,3\cdot 10^{-4}
\end{equation}
With this value of $\epsilon$ there are already no problems with
the radion field.

\section{Localization of matter}
In (\ref{tensor1}) the factor $e^{-2\sigma}$, describing
distribution of matter in the extra dimension, was utilized.
However one can say that matter is spread in the whole bulk rather
than localized on the brane (in analogy to the massless graviton
whose wave function is $\sim e^{2\sigma}$), and it is not evident,
why we make all calculations for $h_{\mu\nu}$ at $y=R$. It seems
that it is more reasonable to consider matter, which is confined
to brane~2 much stronger, since the description of gravity on
brane~2 by $h_{\mu\nu}|_{y=R}$ is more justified in this case. Let
us consider a "toy model"\ with energy-momentum tensor of the form
\begin{equation}\label{tensor-m-1}
T_{\mu\nu}=t_{\mu\nu}(x)\frac{k(1+N)}{e^{2kR+2NkR}-1}\,e^{-2\sigma(1+N)},
\end{equation}
where the function of localization is normalized to unity as well.
Other components of the five-dimensional energy-momentum tensor
are chosen to be
\begin{equation}\label{tensor-m-2}
T_{\mu 4}=0,
\end{equation}
\begin{eqnarray}\label{tensor-m-3}
T_{44}=t(x)\left(k+kN\right)e^{-4\sigma}\left(\frac{c}{2NkR}-\right.\\
\nonumber - \left.\frac{c\epsilon e^{2NkR}}
{e^{2kR+2NkR}-1}-\frac{k}{2Nk\left(e^{2kR+2NkR}-1\right)}\,e^{-2N\sigma}\right).
\end{eqnarray}
Such a complicated form of (\ref{tensor-m-3}) is caused by the
requirements to satisfy the energy conservation law (\ref{econl})
and to get a weak coupling constant for the radion.

The substitution, which allows one to decouple and solve equations
(\ref{mu-4}), (\ref{4-4}) and $\mu\nu$-equation with
(\ref{tensor-m-1}), (\ref{tensor-m-2}) and (\ref{tensor-m-3}), has
the following form:
\begin{eqnarray}\label{subst3}
b_{\mu\nu}=u_{\mu\nu}+\left(\frac{e^{2kR}\left(1-e^{-2NkR}\right)}{4k^2N\epsilon
\left(e^{2kR}-1\right)}-\frac{\left(e^{2kR}-e^{-2NkR}\right)}{8k^2N\epsilon\left(e^{2kR}-1\right)}\,e^{-2\sigma}\right.+\\
\nonumber \left.
+\frac{e^{-2NkR}}{4Nk\left(Nk+2k\right)\epsilon}\,e^{-2\sigma-2N\sigma}\right)
\partial_{\mu}\partial_{\nu}\phi,
\end{eqnarray}
Analogously to what was made in Section~3, we can get
\begin{equation}\label{phi-m}
\Box\phi=\frac{\hat\kappa\epsilon\left(k+kN\right)e^{2NkR}}{3\left(e^{2kR+2NkR}-1\right)}t,
\end{equation}
\begin{equation}\label{alpha-m}
\Box (\alpha_{\mu\nu}-\frac{1}{2}\eta_{\mu\nu}\alpha)
=-\frac{\hat\kappa
k\left(k+Nk\right)\left(1-e^{-2NkR}\right)}{Nk\left(e^{2kR}-1\right)
\left(1-e^{-2kR-2NkR}\right)}t_{\mu\nu}.
\end{equation}
Using (\ref{subst3}) and passing to Galilean coordinates at $y=R$,
in the zero mode approximation we get
\begin{eqnarray}\label{eqbrane-m}
\Box\left(h_{\mu\nu}-\frac{1}{2}\eta_{\mu\nu}h\right)|_{y=R}=\\
\nonumber =-\frac{\hat\kappa
c\left(1+N\right)\left(1-e^{-2NkR}\right)
}{NR\left(1-e^{-2kR-2NkR}\right)}\left(t_{\mu\nu}-\frac{2NkR\epsilon}{3\left(1-e^{-2NkR}\right)}
\left(\eta_{\mu\nu}-\frac{\partial_{\mu}\partial_{\nu}}{\Box}\right)t\right).
\end{eqnarray}
One can see that in the limit $N\to 0$ equation (\ref{eqbrane-m})
passes into (\ref{eqbrane}). The four-dimensional gravitational
constant has the form
\begin{equation}\label{gravconst-m}
G=\hat G \frac{c\left(1+N\right)\left(1-e^{-2NkR}\right)
}{NR\left(1-e^{-2kR-2NkR}\right)}
\end{equation}
(compare with (\ref{BDeq})). For a large $N$ we get
\begin{equation}
G\approx\hat G \frac{c}{R}.
\end{equation}
A choice of relatively small $\epsilon$ makes the contribution of
the radion to be not in contradiction with the present-day
experimental data.

\section{Conclusion and final remarks}
There may arise a question about generation of the new hierarchy
instead of the one solved in the Randall-Sundrum model. Actually,
the factor $\epsilon$ in (\ref{tensor2}) is much smaller than the
factor $e^{2kR}$ in (\ref{tensor2}). But for the energy-momentum
tensor on brane~2 (at $y=R$) we get
\begin{eqnarray}\nonumber
T_{\mu\nu}|_{y=R}=t_{\mu\nu}\frac{e^{2kR}c}{R}\approx k
t_{\mu\nu},\\ \nonumber
T_{44}|_{y=R}=t\frac{e^{4kR}c^{2}}{R}\left(1-4.3\cdot
10^{-4}\right)\approx t\, k^{2}R\left(1-4.3\cdot 10^{-4}\right),
\end{eqnarray}
and the correction $\sim\epsilon$ is not so small in comparison
with the main value $\sim 1$ (analogous calculations can be made
in the case (\ref{tensor-m-1}) and (\ref{tensor-m-3}) too). It
seems to be quite reasonable. One should take into account that
the energy-momentum tensors of the form (\ref{tensor1}),
(\ref{tensor3}), (\ref{tensor2}) and (\ref{tensor-m-1}),
(\ref{tensor-m-2}), (\ref{tensor-m-3}) do not correspond to any
real field action, and must be interpreted as "toy models". There
also arises the question about the origin of such correction
$\sim\epsilon$. Explicitly the term $\sim e^{-4\sigma}$ in
(\ref{tensor2}) and (\ref{tensor-m-3}) corresponds to the
homogeneous solution of equation (\ref{econl}) (i.e. to the case
of $T_{\mu\nu}=0$). Nevertheless such term can be taken into
consideration, since its existence does not evidently contradict
our assumptions. Surely it would be better to derive similar form
of the energy-momentum tensor in a more natural way. For example,
simple examples with scalar and gauge fields in the bulk in the
Randall-Sundrum model were discussed in papers \cite{GW,DHR}. Wave
functions of these fields have their maximum values on brane~2,
and these solutions can be treated as a basis for our choice of
the energy-momentum tensor. But even in the simplest case of the
scalar field only, there arises a nonzero $T_{\mu 4}$-component of
the energy-momentum tensor, which does not satisfy the condition
(\ref{tensor3}). Moreover, the wave functions are different not
only for different fields, but even for different modes. Thus,
formulas (\ref{tensor1}), (\ref{tensor3}), (\ref{tensor2}) and
(\ref{tensor-m-1}), (\ref{tensor-m-2}), (\ref{tensor-m-3}) must be
used as "toy models"\ for classical objects, but not for the
fields (for example, one can choose $t_{00}\sim m\delta (\vec x)$,
$t_{i 0}=0$, $t_{ij}=0$ for a static point-like source).

In this paper we have discussed only one aspect of  the radion
field  problem. Indeed, though the coupling constant of the radion
to matter can be small enough to fit the experimental data, the
size of extra dimension in the RS1 model should be stabilized in
any case. It is well known that in this model gravity alone cannot
stabilize the size of extra dimension due to the  Casimir effect
\cite{GPT}. But if matter is not localized exactly on the brane,
i.e. it has wave functions in the extra dimension, it is possible
that the contribution of matter fields to the  Casimir force
between two branes could  stabilize the size of the extra
dimension. Moreover, even in the case of classical stabilization
with the help of scalar field (such as in \cite{DeWolfe}) the
coupling constant of the massive radion to matter can vary
essentially depending on the profile of matter in the extra
dimension.

There is another question, which may arise in connection with our
choice for matter distribution: why do we choose non-delta-like
profile for matter, whereas the branes have delta-like profile? Of
course, such profile of the branes is an idealization. But one can
recall kink-like solution found in \cite{Kehagias}, which
describes a delta-like brane in the thin brane limit. Thus in the
case under consideration the RS1 setup can be regarded as a
phenomenological model for describing the branes in the case when
the thickness of branes is much smaller than the effective width
of the wave functions of matter. Another advantage of the RS1
background is its relative simplicity, which allowed us to
decouple and solve exactly equations of motion for the radion
field and tensor modes at least in the zero mode approximation and
to show possible effects which can be produced by a non-standard
matter distribution. In a more realistic model the non-zero width
of the branes should be definitely taken into account, but it is
evident that effect analogous to that described above should exist
in this case too.

It should be also noted that the  longitudinal parts of
substitutions (\ref{substitution}), (\ref{substitution1}) and
(\ref{subst3}) contain some terms, which can lead (but it is not
necessarily so) to the strong coupling effect, analogously to what
happens in the DGP model \cite{Luty,Rub}. These terms are pure
gauge from the four-dimensional point of view from the brane and
are not dangerous in the linear order. Nevertheless it is
necessary to consider the non-linear corrections to the equations
of motion to discover the contribution of these longitudinal
terms. But this problem deserves an additional detailed
investigation.

\bigskip
{\large \bf Acknowledgments}
\medskip \\
The authors are grateful to Yu.A.~Kubyshin for valuable
discussions. The work was supported by the RFBR grants 04-02-16476
and 04-02-17448, by the grant UR.02.02.503 of the scientific
program "Universities of Russia", and by the grant NS.1685.2003.2
of the Russian Federal Agency for Science.

\end{document}